\documentclass[conference]{IEEEtran}
\usepackage[T1]{fontenc}
\usepackage{cite}
\usepackage{amsmath,amssymb,amsfonts}
\usepackage{graphicx}
\usepackage{textcomp}
\usepackage{xcolor}
\usepackage{booktabs}
\usepackage{url}

\begin{document}

\title{Predicting Struggling Students in CS1 Programming Using Keystroke-Level Editing Features}

\author{
\IEEEauthorblockN{Yasuyo Kofune}
\IEEEauthorblockA{Graduate School of Science and Technology \\
Nara Institute of Science and Technology \\
Nara, Japan \\
kofune.yasuyo.kv0@is.naist.jp}
 \and
  \IEEEauthorblockN{Kazumasa Shimari}
  \IEEEauthorblockA{
    Faculty of Systems Engineering\\
    Wakayama University\\
    Wakayama, Japan \\
    shimari@wakayama-u.ac.jp
  }
\and
  \IEEEauthorblockN{Kenichi Matsumoto}
  \IEEEauthorblockA{
    Graduate School of Science and Technology\\
    Nara Institute of Science and Technology\\
    Nara, Japan \\
    matumoto@is.naist.jp
  }
}

\maketitle

\begin{abstract}
This paper investigates the feasibility of early detection of struggling students during CS1 programming exercises using keystroke-level logs.
Some students fail to reach a correct solution before the exercise ends, and by the time this becomes apparent from grades or final outcomes, the opportunity for timely instructor support aimed at helping them recover may have passed.
We use data from the CodeBench platform, which records real-time code editing events at the keystroke level, alongside execution and submission logs.
We define two outcome groups: \textit{Breakthrough} (BT) students, whose prior submissions all receive 0\% and whose final submission achieves full credit, and \textit{Fully Stuck} (FS) students, whose submissions all receive 0\% without reaching a correct solution.
To examine this feasibility, we focus on two questions: (RQ1) whether adding keystroke-level editing features improves the prediction of FS students over execution-log features alone, and (RQ2) at which stage BT and FS students can be predicted most accurately.
Experiments on the 2019-1 semester of the CodeBench dataset, comprising 507 students, compare three feature configurations: execution-based features (ExecOnly), CodeMirror-based features (CMOnly), and their combination (Combined).
We evaluate prediction across successive submission-based stages during each exercise.
In the earliest stage, CMOnly outperforms ExecOnly (AUROC 0.654 vs.\ 0.575), and Combined further improves over ExecOnly by $+$0.098 (AUROC 0.674).
Across all configurations, the earliest stage yielded the strongest predictive signal.
These findings indicate that behavioral signals present at the very start of an exercise contain useful clues about whether a student will ultimately solve the problem, and that keystroke-level editing logs provide additional value for early prioritization beyond execution logs alone.
\end{abstract}

\begin{IEEEkeywords}
early prediction, struggling students, keystroke logs, programming education, CS1
\end{IEEEkeywords}

\section{Introduction}

CS1 programming courses consistently exhibit high failure and dropout rates, posing a significant challenge for instructors and institutions worldwide \cite{llanos2023,tabanao2011}.
Timely identification of students who are struggling is essential for enabling effective instructor support; however, in most digital learning environments, the information available to instructors is limited to submission outcomes and final grades---data that only becomes available after a student has finished working on an exercise.

Prior work on early prediction in CS1 courses has addressed this challenge by developing models that predict end-of-course performance at early points in the semester---as early as the first two weeks \cite{pereira2019,pereira2021}, or at weeks 3, 5, and 7 \cite{llanos2023}.
While these approaches have achieved notable predictive accuracy, they rely on features that are only available \textit{after} a student has completed and submitted an exercise---including the number of attempts, total time spent, and assignment grades.
As a result, they are inherently limited to post-submission analysis and cannot inform real-time instructor support during the problem-solving process itself.

The CodeBench platform, used for CS1 programming instruction at the Federal University of Amazonas, Brazil \cite{coelho2024}, is particularly well suited to address this limitation.
In addition to execution and submission logs, CodeBench records keystroke-level editing events through its integrated code editor (CodeMirror).
These logs capture how students write and revise code between submissions---information that is available in real time, as students work on the exercise.
Despite this, prior work using CodeBench has made limited use of CodeMirror logs.
Pereira et al.\ \cite{pereira2019}, \cite{pereira2021} incorporated keystroke-derived features (e.g., keystroke latency, deleted characters, log lines) as session-level aggregate values; however, SHAP analysis in Pereira et al.\ \cite{pereira2021} revealed these features had minimal predictive contribution, with grade-based and submission-based features dominating the model.
The potential of CodeMirror logs as in-exercise behavioral indicators---capturing coding patterns within individual submission intervals before final exercise outcomes are known---thus remains unexplored.
Of particular interest is the earliest stage of an exercise: the period up to and including the first submission, which represents the earliest point at which detection could help instructors prioritize support for students who are likely to remain stuck.

To define the target group without relying on arbitrary thresholds, we use two submission patterns: \textit{Breakthrough} (BT), in which a student receives 0\% on all submissions except the final one, which achieves full credit, and \textit{Fully Stuck} (FS), in which a student receives 0\% on all submissions without reaching a correct solution.
FS students are the primary target of early detection; BT students, who eventually break through on their own, serve as the contrast group.
This BT/FS contrast is intentionally strict: both groups initially experience failure, but only BT students eventually reach a correct solution; thus, the task is not to distinguish successful students from unsuccessful students in general, but to identify students who will remain stuck among those who are already struggling.

Our goal is to examine whether students who will remain fully stuck can be identified as early as possible during an exercise, when instructors still have the greatest opportunity to provide support before the exercise ends.
We operationalize this goal through the following two research questions:

\begin{itemize}
  \item \textbf{RQ1:} Does adding keystroke-level editing features improve the prediction of FS students over execution-log features alone?
  \item \textbf{RQ2:} At which stage of an exercise can BT and FS students be predicted most accurately, and how early can useful predictive signals be observed?
\end{itemize}

To investigate these questions, we divide each exercise session into submission-based segments and evaluate prediction at each stage.
The contributions of this study are as follows:
\begin{enumerate}
  \item We propose a segment-based classification framework that uses CodeMirror and execution logs recorded during an exercise to provide an early prioritization signal for students who may remain stuck, potentially enabling timely instructor support.
  \item (RQ1) We demonstrate that adding keystroke-level features improves prediction over execution logs alone ($+$0.098 AUROC in the initial segment), showing that CodeMirror logs capture behavioral information not reflected in submission outcomes.
  \item (RQ2) We show that the initial segment yielded the strongest segment-wise predictive signal across all feature configurations, indicating that clues about the eventual outcome emerge from the very start of an exercise.
  Notably, detection at the initial segment requires no prior submission history.
\end{enumerate}

The remainder of this paper is organized as follows.
Section~\ref{sec:related} reviews related work.
Section~\ref{sec:methods} describes the dataset, label definitions, and experimental setup.
Section~\ref{sec:results} presents the results.
Section~\ref{sec:discussion} discusses the findings and limitations.
Section~\ref{sec:conclusion} concludes the paper and outlines directions for future work.

\section{Related Work}
\label{sec:related}

Research on early detection of struggling students in CS1 programming courses falls into two broad categories: approaches that rely on outcome-level data (grades, submission counts, time-on-task at the assignment level), and approaches that leverage process-level data recorded during code writing.

\textbf{Outcome-level prediction.}
Several studies have demonstrated that end-of-course performance can be predicted from features accumulated over the first few weeks of a course.
Llanos et al.\ \cite{llanos2023} trained a Gradient Boosting Classifier on grades, delivery time, and attempt counts derived from an online judge platform, achieving F1 scores above 86\% as early as week~3.
Gordon et al.\ \cite{gordon2023} showed that three lightweight engagement metrics---earnestness, struggle, and lab completion---automatically logged by a commercial textbook system suffice to identify at-risk students with 85\% accuracy by week~4 of a CS1 course.
Pereira et al.\ \cite{pereira2021} built an XGBoost model using 20 features from six semesters of CodeBench data, including grades, error quotients, and keystroke-derived aggregates, reaching an AUROC of 0.89; however, SHAP analysis revealed that grade-based and submission-based features dominated the model, with keystroke-derived features contributing minimally.
While effective, these approaches generally rely on accumulated course- or assignment-level outcomes, making them less suitable for intervention during the early stages of an ongoing exercise.

\textbf{Process-level and in-exercise data.}
A smaller body of work examines behavioral signals recorded during the coding process itself.
Using the Error Quotient (EQ) introduced by Jadud \cite{jadud2006}, Tabanao et al.\ \cite{tabanao2011} found significant differences between at-risk and average students and developed regression models that predicted midterm exam scores from BlueJ compilation logs (Adj $R^2 \approx 0.30$), while the accurate identification of individual at-risk students remained challenging.
Gao et al.\ \cite{gao2021} applied differential sequence mining to programming process logs to extract behavioral patterns predictive of students' final course performance, while Dong et al.\ \cite{dong2021} proposed indicators of progress and stagnation from trace logs during problem-solving.
More recently, Pahi et al.\ \cite{pahi2026} collected 20-second code snapshots during in-class exercises and used PCA and K-means clustering to identify three behavioral profiles correlated with exam performance; however, their approach relies on periodic snapshots and aggregate engagement measures rather than fine-grained keystroke-level edit dynamics, and was evaluated on a small sample of graduate students ($n{=}57$) in a monitored classroom setting.

Wang et al.\ \cite{wang2017} applied an LSTM to code-submission sequences from Hour of Code Exercise 18 to predict success on Exercise 19.
Vihavainen et al.\ \cite{vihavainen2013} used weekly aggregates of programming behavior to predict whether students passed a concurrent mathematics course, whereas Jadud \cite{jadud2006} analyzed edit-compile cycles from compiler-triggered code snapshots.
The present study instead extracts keystroke-derived features from individual submission intervals.
These features capture editing behavior before exercise completion and are evaluated at successive stages to predict the final outcome of the same exercise.
Unlike Pereira et al.\ \cite{pereira2019,pereira2021}, who included CodeMirror-derived features only as session-level aggregates, we treat editing behavior within each interval as the primary signal.

\section{Methods}
\label{sec:methods}

\subsection{Dataset}
The CodeBench dataset was used in this study \cite{coelho2024}.
CodeBench is a programming exercise platform developed at the Federal University of Amazonas, Brazil, which records students' activity during Python programming exercises in CS1 courses.
Two types of log data are available for each student--exercise pair: execution/submission logs (executions), and keystroke-level editing logs (CodeMirror), which are generated by CodeMirror, a JavaScript-based code editor integrated into CodeBench.
Grade information is also recorded for each submission.
The 2019-1 semester was selected as a representative pre-COVID-19 semester for this feasibility study; semesters from 2020-ERE through 2020-2 were excluded due to emergency remote education conditions introduced by COVID-19, which may have produced atypical behavioral patterns.
Data from the 2019-1 semester, comprising 507 students, were analyzed.

\subsection{Label Definition}
Two outcome labels were defined based on the submission history of each student--exercise pair.
Only pairs in which all submissions received 0\% were considered, except for BT instances described below.

\begin{itemize}
  \item \textbf{Breakthrough (BT):} A pair with at least one prior 0\% submission followed by a final 100\% submission.
  This pattern represents a student who struggled throughout the exercise but ultimately reached a correct solution.
  A total of 11,999 BT instances were identified in the 2019-1 semester.
  \item \textbf{Fully Stuck (FS):} A pair in which all submissions received 0\% and no correct solution was reached.
  A total of 2,101 FS instances were identified.
\end{itemize}

\begin{figure}[t]
  \centering
  \includegraphics[width=0.85\columnwidth]{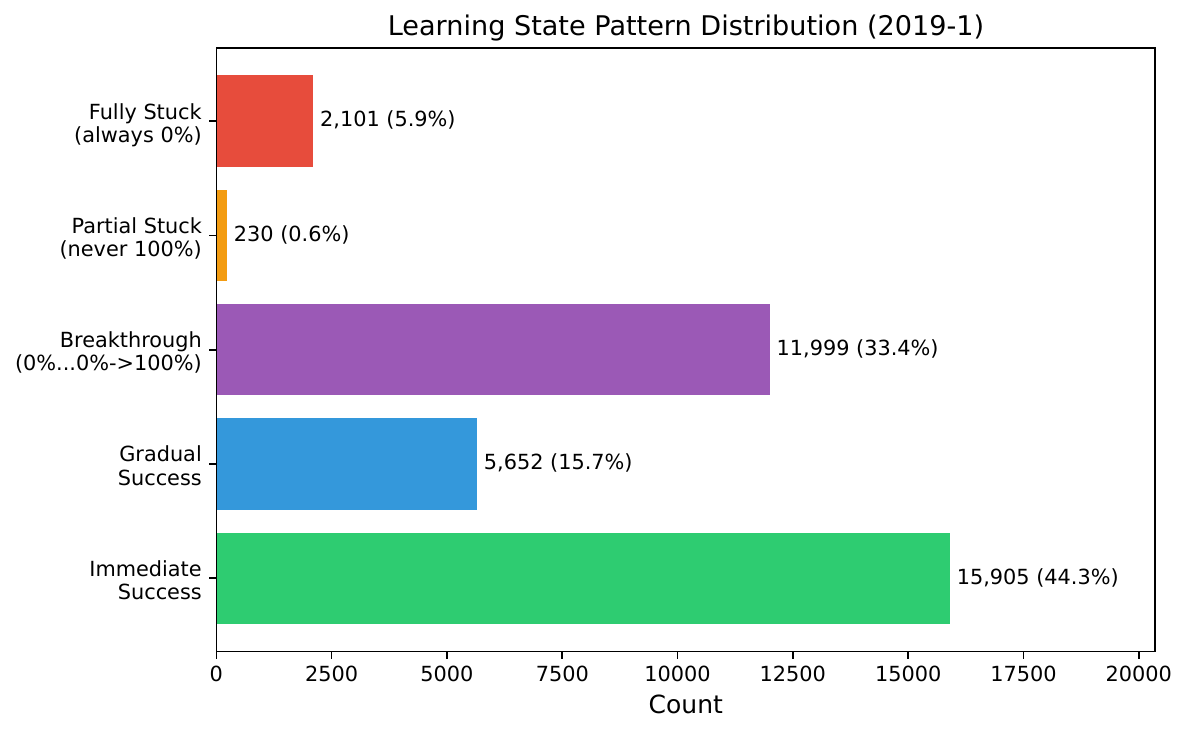}
  \caption{Distribution of submission outcome patterns in the 2019-1 semester (35,887 student--exercise pairs).}
  \label{fig:pattern}
\end{figure}

Pairs in which any submission received partial credit (33\%, 50\%, or 67\%) were excluded.
Figure~\ref{fig:pattern} shows the distribution of all five submission outcome patterns in the 2019-1 semester.
Of the 35,887 pairs, Immediate Success accounted for 15,905 (44.3\%), Breakthrough for 11,999 (33.4\%), Gradual Success for 5,652 (15.7\%), Fully Stuck for 2,101 (5.9\%), and Partial Stuck for 230 (0.6\%).

\subsection{Segment Definition}
A \textit{segment} was defined as the interval between two consecutive submissions.
Segment $k$ spans from the $(k{-}1)$-th submission to the $k$-th submission, where segment~1 spans from the start of the exercise to the first submission.

For BT instances, the final submission (which achieved 100\%) was excluded from analysis, as the editing behavior in that segment reflects solution-oriented activity rather than struggle.
Therefore, the number of segments used for analysis for a BT instance equals the total number of submissions minus one.
For FS instances, all segments were used.

Segments up to $k{=}10$ were analyzed.
Beyond this point, the number of available instances decreased rapidly, making reliable estimation difficult.

Instances for which no editing events were recorded in a given segment were excluded from analysis.
The most common case was at $k{=}1$, where some students submitted without any captured CodeMirror events (e.g., by submitting pre-written code with no in-editor edits).
This exclusion accounts for the difference between the 11,999 total BT pairs and the 10,693 BT instances at $k{=}1$ (similarly, 1,906 of 2,101 FS instances).
At $k{=}1$, the exclusion rate was 10.9\% for BT and 9.3\% for FS, a difference of 1.6 percentage points.
Thus, exclusion frequency was similar between the two outcome groups.

\subsection{Features}
Features were extracted from two sources for each segment.

\textbf{Execution features (ExecOnly)} were derived from the execution/submission log data and captured the outcome of each submission: \texttt{exec\_error\_syntax} (syntax/indentation error; 0/1), \texttt{exec\_error\_runtime} (runtime error; 0/1), \texttt{exec\_error\_logic} (code runs but output is incorrect; 0/1), and \texttt{exec\_sub\_to\_sub\_sec} (elapsed time since previous submission in seconds; 0 for $k{=}1$).

\textbf{CodeMirror features (CMOnly)} were derived from the keystroke-level editing log data and captured the student's coding behavior within the segment: \texttt{n\_edits} (number of edit events), \texttt{deleted\_chars} (deleted characters), \texttt{delete\_ratio} (ratio of deleted characters to total characters changed, i.e., \texttt{deleted\_chars} / (\texttt{deleted\_chars} + \texttt{added\_chars})), \texttt{code\_change} (total characters changed), \texttt{max\_pause\_sec} (maximum inter-keystroke pause duration), \texttt{mean\_pause\_sec} (mean inter-keystroke pause duration), \texttt{duration\_sec} (elapsed time from the first to the last edit event in the segment), and \texttt{idle\_time\_ratio} (proportion of \texttt{duration\_sec} occupied by inter-keystroke pauses shorter than 300\,s; pauses $\geq$300\,s are treated as potential away-from-keyboard intervals and excluded from the numerator).
Idle periods longer than 300\,s were treated as inactivity and excluded when computing editing-duration features.

Three feature configurations were compared: \textbf{ExecOnly} (4 features), \textbf{CMOnly} (8 features), and \textbf{Combined} (12 features).

\subsection{Classification Models and Evaluation}
A Random Forest classifier (RF; \texttt{n\_estimators=200}) and a Logistic Regression classifier (LR; \texttt{max\_iter=1000}, L2 regularization, \texttt{C=1.0}) were trained separately for each segment $k$ and feature configuration.
Classification performance was evaluated using student-level GroupKFold cross-validation (5 folds), ensuring that all exercise records from the same student were assigned to the same fold to prevent data leakage.
Although each student contributed multiple exercise records (mean $\approx$ 27 exercises/student), 85.7\% of students exhibited both BT and FS patterns across different exercises, reducing the likelihood that classification performance can be explained solely by student identity.
AUROC was selected as the primary metric because it is threshold-independent and robust to class imbalance.
AUROC values were computed from pooled out-of-fold (OOF) predictions across the five GroupKFold splits, yielding a single consistent estimate over the full dataset.
To assess the statistical reliability of AUROC differences, paired bootstrap confidence intervals ($N{=}2{,}000$) were computed for the AUROC differences between CMOnly and ExecOnly, and between Combined and ExecOnly, at each segment $k$ by resampling students with replacement and including all prediction instances for each resampled student (group bootstrap); the 95\% CI was estimated as the 2.5th and 97.5th percentiles of the bootstrap distribution.
As a robustness check, recomputing the intervals with instance-level resampling yielded nearly identical intervals.
No class weighting or resampling was applied.
In addition to AUROC, PR-AUC (average precision) was computed to evaluate FS detection under class imbalance, treating FS as the positive class; the FS prevalence at $k{=}1$ (0.151) serves as the random-classifier baseline for PR-AUC. Each model is trained and evaluated using only the features extracted within that segment; this segment-wise design compares prediction accuracy at each stage independently, rather than accumulating information across all preceding segments.

As a supplementary deployment-oriented analysis, we also evaluated an FS-vs-rest setting at $k{=}1$, in which FS was treated as the positive class and all other outcome patterns were treated as negative.
After applying the same exclusion criteria as in the main analysis, including the exclusion of pairs without CodeMirror events at $k{=}1$, this setting contained 31,790 student--exercise pairs across the five outcome categories, with an FS prevalence of 0.060.

\section{Results}
\label{sec:results}

\subsection{Comparison of Feature Configurations (RQ1)}

\begin{figure}[t]
  \centering
  \includegraphics[width=\columnwidth]{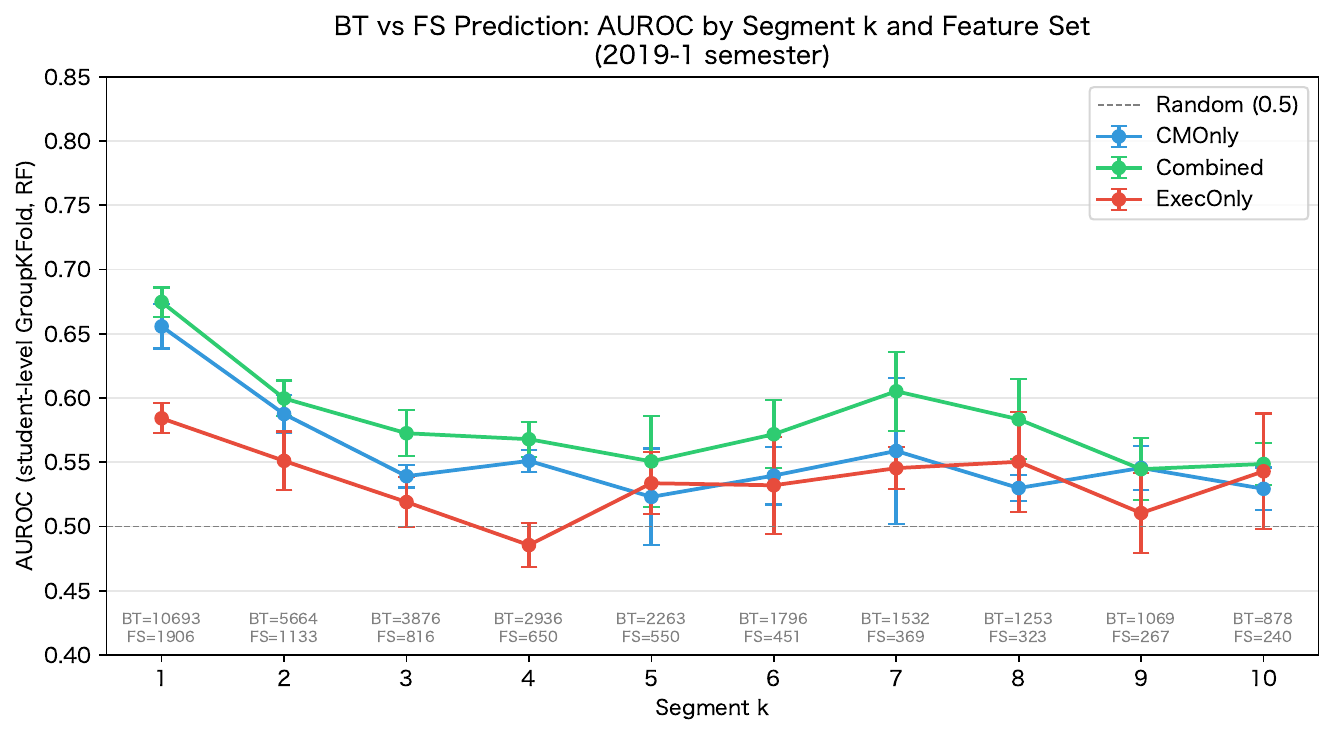}
  \caption{AUROC by submission-based segment and feature configuration.}
  \label{fig:auc}
\end{figure}

Figure~\ref{fig:auc} presents the AUROC scores for the three feature configurations across segments ($k{=}1$--10).

\begin{figure}[t]
  \centering
  \includegraphics[width=\columnwidth]{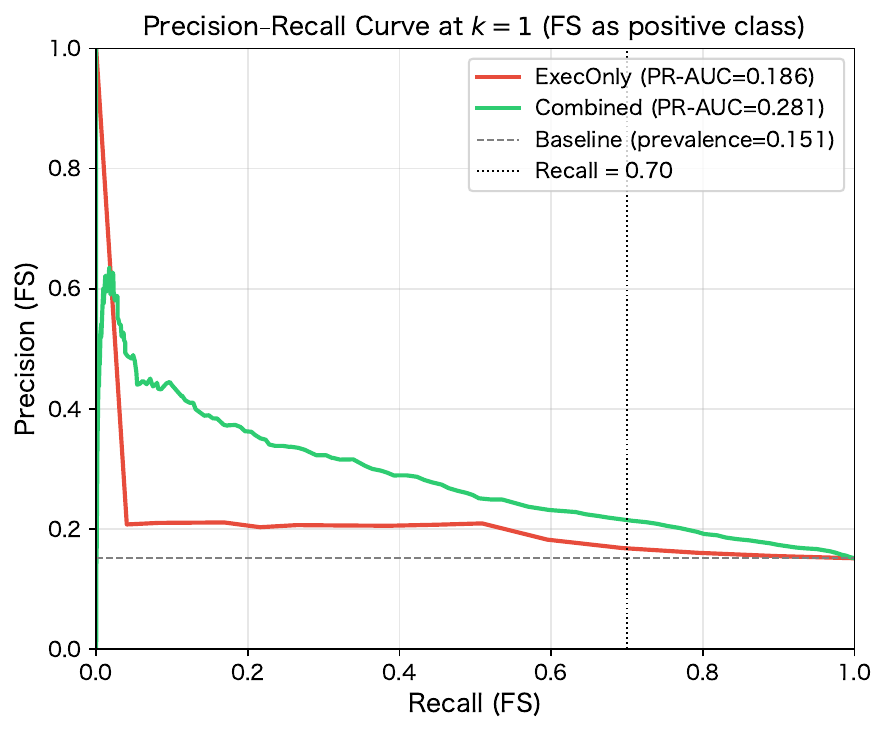}
  \caption{Precision--recall curves for the first segment.}
  \label{fig:pr}
\end{figure}

Three findings address RQ1.
First, CMOnly outperformed ExecOnly at $k{=}1$ (0.654 vs.\ 0.575), with a bootstrap 95\% confidence interval for the AUROC difference of [$+$0.056, $+$0.104], indicating that keystroke-level features derived from CodeMirror logs are more predictive than execution-based features alone at the earliest stage of an exercise.
Second, Combined outperformed ExecOnly with statistically reliable improvements in the early segments ($k{=}1$--4), as confirmed by paired bootstrap confidence intervals ($N{=}2{,}000$); the 95\% CI of the AUROC difference at $k{=}1$ was [$+$0.078, $+$0.119], with the largest gain of $+$0.098 (AUROC 0.674).
Later segments ($k{\geq}5$) showed a less consistent pattern, with confidence intervals spanning zero at most points.
Third, at $k{=}4$, ExecOnly fell near chance level (AUROC 0.484) while Combined maintained 0.568 (95\% CI: [$+$0.052, $+$0.116]), suggesting that CodeMirror features preserve discriminability even when execution-based signals become uninformative.
The same ordering was observed in PR-AUC at $k{=}1$: Combined achieved the highest PR-AUC (0.281), followed by CMOnly (0.256) and ExecOnly (0.186), all exceeding the FS prevalence baseline of 0.151.
Figure~\ref{fig:pr} shows the Precision--Recall curves for ExecOnly and Combined at $k{=}1$.
At the operating point where FS recall was set to 0.70, the Combined model achieved a precision of 0.214, indicating that approximately two of every ten flagged student--exercise pairs would be FS cases.

\subsection{Prediction Performance Across Segments (RQ2)}

Figure~\ref{fig:auc} also reveals how prediction performance changes as a function of segment index.
The highest AUROC was achieved at $k{=}1$ across all configurations, with the Combined model reaching 0.674.
Combined and CMOnly showed a general declining trend with increasing $k$ (from 0.674 to 0.549 and from 0.654 to 0.529, respectively).
ExecOnly, by contrast, exhibited a non-monotonic pattern: it dipped near chance level at $k{=}4$ (0.484) before recovering to 0.543 at $k{=}10$, with no clear downward trend.
These results indicate that the first segment contained the strongest segment-wise predictive signal.

\subsection{Feature Importance}

To understand what drives the improvement observed in RQ1, feature importance was examined descriptively using the Random Forest classifier retrained on the full dataset for each segment (CMOnly configuration).
Table~\ref{tab:importance} reports Gini importance values for $k{=}1$, 3, and~6 as representative segments.

\begin{table}[t]
\caption{Gini feature importance for the CMOnly configuration.}
\label{tab:importance}
\centering
\begin{tabular}{lrrr}
\toprule
Feature & $k{=}1$ & $k{=}3$ & $k{=}6$ \\
\midrule
\texttt{duration\_sec}     & \textbf{.149} & \textbf{.179} & .170 \\
\texttt{code\_change}      & .139 & .108 & .118 \\
\texttt{max\_pause\_sec}   & .131 & .171 & \textbf{.171} \\
\texttt{mean\_pause\_sec}  & .128 & .174 & .165 \\
\texttt{delete\_ratio}     & .127 & .111 & .113 \\
\texttt{n\_edits}          & .127 & .105 & .107 \\
\texttt{deleted\_chars}    & .108 & .093 & .100 \\
\texttt{idle\_time\_ratio} & .091 & .059 & .057 \\
\bottomrule
\end{tabular}
\end{table}

Temporal features, particularly \texttt{duration\_sec} and \texttt{max\_pause\_sec}, consistently ranked among the top three features across the examined segments.
In addition, \texttt{mean\_pause\_sec} ranked within the top three from $k{=}3$ onward.
Notably, \texttt{code\_change} ranked second at $k{=}1$ but declined thereafter, suggesting that sheer volume of editing activity is particularly informative before the first submission but becomes less distinguishing in later segments.
This pattern indicates that temporal aspects of student behavior---how long students work and how they pause---are more informative for predicting the BT/FS outcome than the quantity of individual edit operations.

Additional baselines at $k{=}1$ showed that a \texttt{duration\_sec}-only LR model achieved only near-chance performance (AUROC 0.473), despite \texttt{duration\_sec} being the highest-ranked RF feature.
In contrast, Combined LR achieved 0.653, close to Combined RF (0.674), indicating that the early predictive signal arises from multivariate behavioral features rather than editing duration alone; nonlinear modeling provided a modest additional gain.

\textbf{Supplementary FS-vs-rest analysis.}
In the FS-vs-rest setting, the Combined model achieved an AUROC of 0.839 and a PR-AUC of 0.224, corresponding to 3.74 times the random-classifier baseline (FS prevalence\,=\,0.060).
The higher AUROC compared with the main BT/FS analysis (0.674) is partly attributable to the inclusion of outcome patterns that are behaviorally distant from FS, such as Immediate Success, making the classification task easier.
These results provide supplementary evidence that FS-specific behavioral signals are detectable in a broader setting, while the main BT/FS analysis remains the stricter evaluation against the hardest contrast group.

\section{Discussion}
\label{sec:discussion}

\subsection{Interpretation of Key Findings}

The most consistent pattern across all feature configurations is that prediction performance was highest at the initial segment and generally declined as $k$ increased.
This may reflect a selection effect: at later segments, only students who have already failed multiple times remain in the analysis, and behavioral differences between BT and FS students may become less pronounced.
The advantage of CMOnly over ExecOnly at the initial segment suggests that keystroke-level editing patterns capture aspects of the problem-solving process not reflected in submission outcomes alone---specifically, how students engage with code before submitting, rather than what errors they produce.
The first segment may capture differences in how students interpret the problem and plan a solution.
After a failed submission, both groups may make similar incremental trial-and-error edits.
This would reduce the behavioral contrast between them.
The smaller number of instances at larger $k$ also increases sampling uncertainty in the corresponding AUROC estimates.

\subsection{Relation to Prior Work}

Prior work has mainly predicted course- or assignment-level outcomes after submissions were completed \cite{pereira2019,pereira2021,llanos2023}.
In contrast, the present results show that keystroke-level behavior observed during an exercise provides above-chance signals for distinguishing BT and FS students, especially at the earliest stage.
Although the AUROC values obtained (up to 0.674) are lower than the F1 scores reported in prior work, such a comparison is not straightforward: predicting outcomes from early in-exercise behavioral signals is a fundamentally harder task than summarizing completed submissions.

\subsection{Limitations}

\textbf{Generalizability.}
The analysis is based on data from a single semester (2019-1) at one institution, and exercise difficulty was not controlled for.
More difficult exercises may produce more FS instances and could confound the relationship between editing features and BT/FS outcomes.
Generalizability to other semesters, institutions, or programming languages remains to be verified. The evaluation also does not test generalization to unseen exercises; future work should examine exercise-level splits or models that account for exercise difficulty.

\textbf{Generative AI-assisted coding.}
The 2019-1 data predate the widespread use of generative AI coding assistants.
These tools may change editing behavior, for example when students insert generated code and make few subsequent edits.
The predictive value of keystroke-derived features in recent cohorts has not been tested.

\textbf{Construct validity of the FS label.}
The FS group may include both persistent struggle and early disengagement.
In the 2019-1 semester, 25.4\% of FS instances had only one submission, although most FS instances involved multiple submissions (median: 3; mean: 8.22).
This raises a construct-validity concern that the model may partly capture early disengagement rather than struggle alone, especially because editing duration was the most important feature.
As a sensitivity analysis, we repeated the $k{=}1$ evaluation after excluding FS instances with only one submission; the Combined AUROC decreased from 0.674 to 0.650, suggesting that early predictive signal remains even among students who made multiple unsuccessful submissions.

\textbf{Operational deployment.}
The class ratio between BT and FS instances is approximately 5.6:1.
When the threshold was set to achieve FS recall of 0.70, precision was 0.214, indicating a substantial false-positive rate.
The proposed model should therefore be viewed as an early prioritization signal rather than a standalone intervention decision tool; operational thresholds require consideration of classroom constraints and instructor capacity.

\textbf{Modeling scope.}
Each segment was modeled independently, so the models did not use sequential dependencies across segments.
Recurrent or attention-based architectures could incorporate these dependencies.
The BT/FS contrast is intentionally strict, focusing on students with clear behavioral endpoints; performance against broader outcome patterns is partially addressed in the supplementary FS-vs-rest experiment.
The maximum AUROC of 0.674, while above chance, suggests that fine-grained behavioral features alone are insufficient for high-confidence individual prediction.

Despite these limitations, the study provides initial evidence that fine-grained editing behavior can serve as an early prioritization signal for students who may remain stuck during programming exercises.
To support reproducibility, the preprocessing and analysis scripts, information on library versions, and instructions for reproducing the results are available at \url{https://doi.org/10.5281/zenodo.20695742}. Access to the CodeBench dataset is subject to the dataset provider's terms of use.

\section{Conclusion}
\label{sec:conclusion}

This study investigated whether keystroke-level CodeMirror logs can provide early signals for prioritizing students who may remain stuck during a programming exercise.
Using data from the 2019-1 semester of the CodeBench platform, we defined two outcome groups---Breakthrough (BT) and Fully Stuck (FS)---and extracted features from execution logs and CodeMirror editing logs for each submission interval (segment).

Experiments comparing three feature configurations (ExecOnly, CMOnly, and Combined) yielded two main findings.
First, regarding RQ1, adding CodeMirror-based features improved prediction performance over execution logs alone ($+$0.098 AUROC at $k{=}1$), showing that keystroke-level editing logs capture behavioral information not reflected in submission outcomes.
Second, regarding RQ2, the first segment ($k{=}1$) consistently achieved the highest prediction performance across all configurations (AUROC 0.674 for Combined), suggesting that behavior in the first segment contains above-chance information about eventual problem-solving outcomes.

Overall, these findings provide preliminary empirical evidence for the feasibility of in-exercise struggle detection using fine-grained editing logs, and offer a basis for future research toward early prioritization and support systems, pending further improvements in predictive precision.

Future work should build on the finding that the earliest submission-based segment contains useful predictive signal.
One direction is to investigate whether such early predictions can be translated into instructional support during this early stage, such as targeted hints, instructor check-ins, or prioritization of students for assistance.
Another direction is to move beyond prediction and characterize the behavioral differences between students who eventually recover and those who remain stuck, which may help explain what the model is detecting and what types of support are most appropriate.
Developing such support systems will also require operational criteria that reflect classroom constraints, including instructor capacity and the acceptable cost of false positives.

\section*{Acknowledgment}
This work has been supported by JSPS KAKENHI No. JP26K21197 and the Support Center for Advanced Telecommunications Technology Research.

\bibliographystyle{IEEEtran}
\bibliography{references}

@article{coelho2024,
  author  = {Fl{\'a}vio J. M. Coelho and Elaine H. T. Oliveira and Filipe D. Pereira
             and David B. F. Oliveira and Leandro S. G. Carvalho and Eduardo J. P. Souto
             and Marcela Pessoa and Rafaela Melo and Marcos A. P. de Lima
             and Fab{\'i}ola G. Nakamura},
  title   = {Learning Analytics in Introductory Programming Courses: {A} Showcase
             from the {Federal University of Amazonas}},
  journal = {Revista Brasileira de Inform{\'a}tica na Educa{\c{c}}{\~a}o},
  volume  = {31},
  pages   = {1089--1127},
  year    = {2023},
  doi     = {10.5753/rbie.2023.3334}
}

@inproceedings{dong2021,
  author    = {Yihuan Dong and Samiha Marwan and Preya Shabrina and Thomas Price
               and Tiffany Barnes},
  title     = {Using Student Trace Logs to Determine Meaningful Progress and Struggle
               During Programming Problem Solving},
  booktitle = {Proc. 14th Int. Conf. Educational Data Mining ({EDM})},
  pages     = {439--445},
  year      = {2021}
}

@inproceedings{gao2021,
  author    = {Ge Gao and Samiha Marwan and Thomas W. Price},
  title     = {Early Performance Prediction Using Interpretable Patterns in
               Programming Process Data},
  booktitle = {Proc. 52nd {ACM} Technical Symp. Computer Science Education ({SIGCSE})},
  pages     = {342--348},
  doi       = {10.1145/3408877.3432439},
  year      = {2021}
}

@inproceedings{gordon2023,
  author    = {Chelsea Gordon and Stanley Zhao and Frank Vahid},
  title     = {Ultra-Lightweight Early Prediction of At-Risk Students in {CS1}},
  booktitle = {Proc. 54th {ACM} Technical Symp. Computer Science Education {V}. 1 ({SIGCSE})},
  pages     = {764--770},
  doi       = {10.1145/3545945.3569764},
  year      = {2023}
}

@inproceedings{jadud2006,
  author    = {Matthew C. Jadud},
  title     = {Methods and Tools for Exploring Novice Compilation Behaviour},
  booktitle = {Proc. 2nd Int. Workshop on Computing Education Research ({ICER})},
  pages     = {73--84},
  doi       = {10.1145/1151588.1151600},
  year      = {2006}
}

@article{llanos2023,
  author  = {Jose Llanos and V{\'i}ctor A. Bucheli and Felipe Restrepo-Calle},
  title   = {Early Prediction of Student Performance in {CS1} Programming Courses},
  journal = {{PeerJ} Computer Science},
  volume  = {9},
  pages   = {e1655},
  year    = {2023},
  doi     = {10.7717/peerj-cs.1655}
}

@inproceedings{pahi2026,
  author    = {Kritish Pahi and Vinhthuy Phan},
  title     = {Using In-Class Exercise Data for Early Support of Struggling Students},
  booktitle = {Proc. 57th {ACM} Technical Symp. Computer Science Education {V}. 1 ({SIGCSE TS})},
  pages     = {797--803},
  doi       = {10.1145/3770762.3772630},
  year      = {2026}
}

@inproceedings{pereira2019,
  author    = {Filipe Dwan Pereira and Elaine H. T. Oliveira and David Fernandes
               and Alexandra Cristea},
  title     = {Early Performance Prediction for {CS1} Course Students Using a
               Combination of Machine Learning and an Evolutionary Algorithm},
  booktitle = {Proc. 19th {IEEE} Int. Conf. Advanced Learning Technologies ({ICALT})},
  pages     = {183--184},
  doi       = {10.1109/ICALT.2019.00066},
  year      = {2019}
}

@article{pereira2021,
  author  = {Filipe Dwan Pereira and Samuel C. Fonseca and Elaine H. T. Oliveira
             and Alexandra I. Cristea and Henrik Bellh{\"a}user and Luiz Rodrigues
             and David B. F. Oliveira and Seiji Isotani and Leandro S. G. Carvalho},
  title   = {Explaining Individual and Collective Programming Students' Behavior
             by Interpreting a Black-Box Predictive Model},
  journal = {{IEEE} Access},
  volume  = {9},
  pages   = {117097--117119},
  year    = {2021},
  doi     = {10.1109/ACCESS.2021.3105956}
}

@inproceedings{tabanao2011,
  author    = {Emily S. Tabanao and Ma. Mercedes T. Rodrigo and Matthew C. Jadud},
  title     = {Predicting At-Risk Novice {Java} Programmers Through the Analysis
               of Online Protocols},
  booktitle = {Proc. 7th Int. Workshop on Computing Education Research ({ICER})},
  pages     = {85--92},
  doi       = {10.1145/2016911.2016930},
  year      = {2011}
}

@inproceedings{vihavainen2013,
  author    = {Arto Vihavainen and Matti Luukkainen and Jaakko Kurhila},
  title     = {Using Students' Programming Behavior to Predict Success in an
               Introductory Mathematics Course},
  booktitle = {Proc. 6th Int. Conf. Educational Data Mining ({EDM})},
  pages     = {300--303},
  year      = {2013}
}

@inproceedings{wang2017,
  author    = {Lisa Wang and Angela Sy and Larry Liu and Chris Piech},
  title     = {Deep Knowledge Tracing on Programming Exercises},
  booktitle = {Proc. 4th {ACM} Conf. Learning @ Scale ({L@S})},
  pages     = {201--204},
  doi       = {10.1145/3051457.3053985},
  year      = {2017}
}

\end{document}